\documentclass[letter, a4paper]{IEEEtran}
\usepackage{amsmath,amsfonts}
\usepackage{algorithmic}
\usepackage{algorithm}
\usepackage{array}
\usepackage{textcomp}
\usepackage{stfloats}
\usepackage{url}
\usepackage{verbatim}
\usepackage{graphicx}
\usepackage{cite}
\usepackage{float} 
\usepackage{authblk}
\usepackage{setspace}
\usepackage{subfigure}

\ifCLASSOPTIONcompsoc
\usepackage[caption=false,font=normalsize,labelfont=sf,textfont=sf]{\part{title}subfig}
\else
\usepackage{subfigure}
\fi

\usepackage{graphicx,amsmath,amssymb,amsfonts}
\allowdisplaybreaks[3]
\usepackage{caption} 
\usepackage[colorlinks,
linkcolor=blue,
anchorcolor=blue,
citecolor=green]{hyperref}

\usepackage{xcolor}
\usepackage[T1]{fontenc}   

\newtheorem{theorem}{\textbf{Theorem}}

\newtheorem{lemma}{\textbf{Lemma}}

\newtheorem{corollary}{\textbf{Corollary}}

\makeatletter

\newcommand{\Rmnum}[1]{\expandafter\@slowromancap\romannumeral #1@}
\newcommand{\tabincell}[2]{\begin{tabular}{@{}#1@{}}#2\end{tabular}}
\makeatother

\begin{document}
	
	\title{Enhancing NOMA Networks via Reconfigurable Multi-Functional Surface}
	
	\author{
		Ailing~Zheng,
		Wanli~Ni, 
		Wen~Wang,
		and Hui~Tian
		\vspace{-8 mm}   
		\thanks{This letter was supported by the Natural Science Foundation of Shandong Province under Grant No. ZR2021LZH010. The associate editor coordinating the review of this letter and approving it for publication was Lina Bariah. \emph{(Corresponding author: Hui Tian.)}
		}
		\thanks{A. Zheng, W. Ni, W. Wang, and H. Tian are with the State Key Laboratory of Networking and Switching Technology, Beijing University of Posts and Telecommunications, Beijing 100876, China (e-mail: \{ailing.zheng, charleswall, wen.wang, tianhui\}@bupt.edu.cn).}
	}
	
	\maketitle
	
	\begin{abstract}
		By flexibly manipulating the radio propagation environment, reconfigurable intelligent surface (RIS) is a promising technique for future wireless communications. However, the single-side coverage and double-fading attenuation faced by conventional RISs largely restrict their applications.
		To address this issue, we propose a novel concept of multi-functional RIS (MF-RIS), which provides reflection, transmission, and amplification simultaneously for the incident signal. With the aim of enhancing the performance of a non-orthogonal multiple-access (NOMA) downlink multiuser network, we deploy an MF-RIS to maximize the sum rate by jointly optimizing the active beamforming and MF-RIS coefficients. Then, an alternating optimization algorithm is proposed to solve the formulated non-convex problem by exploiting successive convex approximation and penalty-based method. Numerical results show that the proposed MF-RIS outperforms conventional RISs under different settings.
	\end{abstract}
	
	\vspace{-2 mm} 
	\section{Introduction}	\label{Introduction}
	Compared to orthogonal multiple access (OMA), non-orthogonal multiple access (NOMA) is capable of achieving high spectrum efficiency and massive connectivity \cite{LiuNOMA2017}. Prior investigations have shown that the differences between users' channel conditions can be exploited to enhance NOMA performance \cite{ElhattabNOMA}.
	However, users in large-scale networks may have poor or similar channel conditions, which hinders the application of successive interference cancellation (SIC) and the effective implementation of NOMA.
	Therefore, adjusting channel conditions and enhancing channel diversity are able to release the potential of NOMA in practical networks.
	
	Recently, with the ability to reshape the wireless propagation environment, reconfigurable intelligent surface (RIS) has emerged as a key technique to improve the performance of NOMA networks \cite{Liu2021RIS}.   
	By properly designing the reflection coefficients, RIS is able to smartly change the combined channels to enhance the differences among users, thus boosting the performance of NOMA in large-scale networks. 
	Initial investigations on RIS-aided NOMA networks in \cite{Liu2021RIS, SenaRIS-NOMA2020,DingRIS-NOMA2020, ZhengRIS-NOMA2020,Ni2021RIS} had verified the superiority of the integration of NOMA and RIS.
	Specifically, the authors of \cite{Liu2021RIS} and  \cite{SenaRIS-NOMA2020} performed comprehensive discussions of the main challenges and futuristic use cases regarding RIS-aided NOMA networks. Moreover, the works in \cite{DingRIS-NOMA2020,ZhengRIS-NOMA2020,Ni2021RIS} demonstrated the benefits brought by RISs to achieve performance trade-off among multiple NOMA users through smartly adjusting the decoding order.
	However, the existing literature on RIS-aided NOMA networks mostly uses single functional RIS (SF-RIS) that only supports signal reflection or transmission/refraction. 
	This implies that only users located in a single side can be served by the SF-RIS if no additional operations are performed.
	
	To overcome this limitation, the authors of \cite{Wang2021STARRIS} proposed the concept of dual-functional RIS (DF-RIS). Unlike SF-RIS, DF-RIS refers to the reconfigurable dual-functional surface that can conduct signal reflection and transmission simultaneously, such as simultaneous transmitting and reflecting RIS (STAR-RIS)\cite{Wu2021STARRIS} and intelligent omni-surface (IOS) \cite{WangIOSrate}.  
	Specifically, the coverage characterization of STAR-RIS-aided NOMA networks was investigated in \cite{Wu2021STARRIS} by studying a coverage range maximization problem.
	The authors of \cite{WangIOSrate} considered the average rate maximization problem in an IOS-aided NOMA networks with spatially correlated channels.
	Furthermore, the effective capacity and secrecy outage probability of STAR-RIS-aided NOMA networks were derived in \cite{Liu2022WCL} and \cite{Li2022TVT}, respectively.
	However, although the effective coverage can be enhanced by the existing DF-RIS, the signals relayed by the DF-RIS still suffer from channel fading twice due to the features of cascaded channels.
	This double-fading effect inevitably deteriorates the achievable performance of passive RIS-assisted wireless networks.
	Therefore, it is necessary to design new RIS architectures to mitigate the double-fading attenuation problem faced by the existing RISs.
	
	In this letter, a novel multi-functional RIS (MF-RIS) is proposed to address the issues aforementioned. Specifically, the proposed MF-RIS can not only divide the incident signal into transmission and reflection two parts based on the field equivalence principle, but also amplify the outgoing signal with the help of active loads. Thus, the MF-RIS is able to facilitate a full-space coverage and overcome the double-fading issue.
	Then, we investigate a sum rate maximization problem in an MF-RIS-aided NOMA network. Compared to the existing problems formulated in \cite{Wang2021STARRIS} and \cite{Wu2021STARRIS}, 
	the newly introduced MF-RIS constraints and highly coupled variables make the performance optimization more complicated. The main contributions of this letter are summarized as follows:
	1) We propose a new concept of MF-RIS by integrating the surface electric and magnetic impedances, and power amplifier into each element so that the incident signal can be reflected, refracted, and amplified simultaneously.  
	2) We formulate a non-convex optimization problem to maximize the throughout of an MF-RIS-aided NOMA network, where the MF-RIS is deployed to constructively enhance the channel condition by flexibly adjusting the radio propagation environment.
	3) To solve the formulated non-convex problem, we propose an efficient iterative algorithm by alternatively optimizing the active beamforming and MF-RIS coefficients based on the penalty-based method and successive convex approximation (SCA).
	4) Simulation results show that the proposed MF-RIS-aided NOMA network can provide up to about 59\% sum rate gain than the SF-RIS, and the MF-RIS prefers to be deployed at the user side for better performance.
	
	\vspace{-4 mm}
	\section{System Model and Problem Formulation}	\label{System Model}
	\begin{figure*}[t]
		\centering
		\includegraphics[width=6 in]{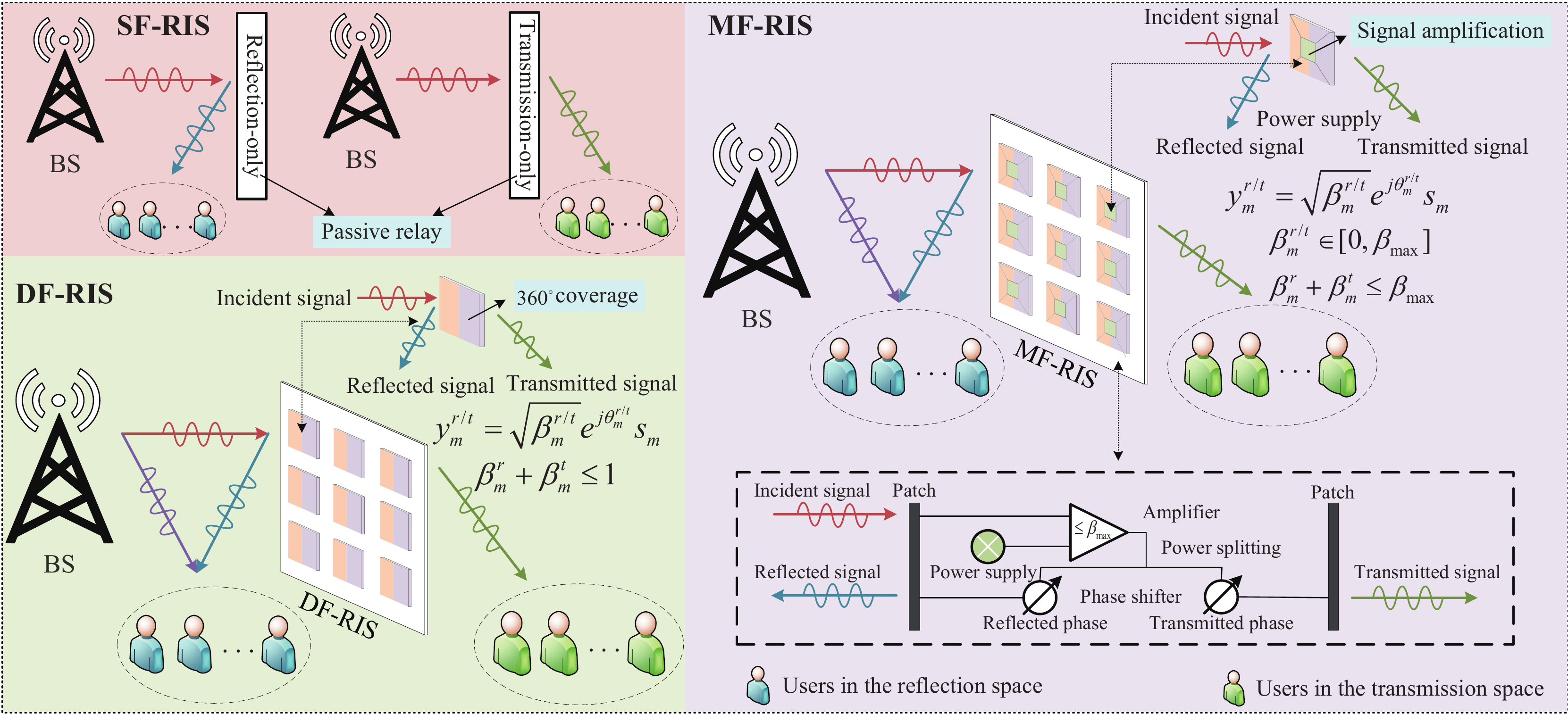}
		\caption{Conventional RIS vs. the proposed MF-RIS-aided NOMA networks.}
		\vspace{-15 pt}
		\label{system model}
	\end{figure*} 
	
	\subsection{System Model} 
	We consider an MF-RIS-aided NOMA downlink network, where an $N$-antenna BS communicates with $K$ single-antenna users with the aid of an MF-RIS comprising $M$ elements,
	as shown in the right of Fig. {\ref{system model}}.
	The sets of elements and users are denoted by $\mathcal{M}=\{1,2,\ldots,M\}$ and $\mathcal{K}=\{1,2,\ldots,K\}$, respectively. 	
	The channels of BS-user, BS-RIS, and RIS-user are denoted by $\mathbf{h}_k\in\mathbb{C}^{N\times1}$, $\mathbf{H}\in\mathbb{C}^{M\times N}$, and $\mathbf{g}_k\in\mathbb{C}^{M\times1}$, respectively. Furthermore, we define $\mathbf{u}_p=[\sqrt{\beta_1^p}e^{j\theta_1^p},~\sqrt{\beta_2^p}e^{j\theta_2^p},\ldots, ~\sqrt{\beta_M^p}e^{j\theta_M^p}]^{\mathrm{T}} \in\mathbb{C}^{M\times1}$ as the transmission $(p=t)$ or reflection $(p=r)$ beamforming vector, where $p \in \{t,r\}$ denotes the transmission and reflection spaces, $\beta_m^p \in [0,\beta_{\max}]$ and $\theta_m^p \in [0,2\pi)$ represent the amplitude and the phase shift response of the $m$-th element, respectively, with the maximum amplification factor $\beta_{\max} \ge 1$. Due to the law of energy conservation, we have $\beta_m^r+\beta_m^t \le \beta_{\rm{max}}$. If user $k$ is located at the reflection space, the diagonal matrix of the MF-RIS for user $k$ is given by $\boldsymbol{\Theta}_k=\mathrm{diag}(\mathbf{u}_r)$; otherwise $\boldsymbol{\Theta}_k=\mathrm{diag}(\mathbf{u}_t)$.
	
	We assume that the perfect channel state information (CSI) of all channels is available at the BS.
	Then the signal received at user $k$ is expressed as
	\setlength{\abovedisplayskip}{3pt}
	\setlength{\belowdisplayskip}{3pt}
	\begin{eqnarray}
		\!\!y_k=(\mathbf{h}_k^{\rm H}+\mathbf{g}_k^{\rm H}\boldsymbol{\Theta}_k\mathbf{H})\mathbf{x}+\mathbf{g}_k^{\rm H}\boldsymbol{\Theta}_k\mathbf{n}_s+n_k,~\forall k,
	\end{eqnarray} 
	where $\mathbf{x}=\sum\nolimits_{k}\mathbf{w}_ks_k$ denotes the transmit signal, $\mathbf{w}_k$ and $s_k\in\mathcal{CN}(0,1)$ represent the transmit precoder and the information symbol for user $k$, respectively. $\mathbf{n}_s\in \mathcal{CN}(\mathbf{0},\sigma_s^2 \mathbf{I}_M)$  denotes the dynamic noise at the MF-RIS with each element's noise power $\sigma_s^2$, and $n_{k} \in \mathcal{CN}(0,\sigma_k^2)$ denotes the additive white Gaussian noise at user $k$ with power $\sigma_k^2$.
	
	By employing SIC, the strong user can mitigate the interference from weak users to improve the signal-to-interference-plus-noise ratio.
	Similar to \cite{SenaRIS-NOMA2020,DingRIS-NOMA2020,ZhengRIS-NOMA2020}, 	
	with the assistance of RIS to flexibly adjust channel conditions of multiple users, we assume that users' indexes are ranked in an increasing order with respect to their channel gains, i.e.,
	\begin{equation}
		\setlength{\abovedisplayskip}{3pt}
		\setlength{\belowdisplayskip}{3pt}	
		\label{channel gain}
		\Vert \widehat{\mathbf{h}}_1\Vert ^2 \le \Vert \widehat{\mathbf{h}}_2\Vert ^2 \le \cdots \le \Vert \widehat{\mathbf{h}}_K\Vert ^2,
	\end{equation}
	where $\widehat{\mathbf{h}}_k=\mathbf{h}_k^{\rm H}+\mathbf{g}_k^{\rm H}\boldsymbol{\Theta}_k\mathbf{H}$ is the equivalent combined channel. 
	
	For the fixed decoding order, the corresponding achievable sum rate of user $k$ is given by $R_k=\log_2(1+\gamma_k)$, where $\gamma_k$ can be obtained by 
	\begin{eqnarray}
		\setlength{\abovedisplayskip}{3pt}
		\setlength{\belowdisplayskip}{3pt}
		\label{SINR}
		\gamma_k=\frac{\vert \widehat{\mathbf{h}}_k\mathbf{w}_k\vert ^2}
		{\sum_{i=k+1}^{K} (\vert \widehat{\mathbf{h}}_{k}\mathbf{w}_{i} \vert ^2)+\vert \mathbf{g}_k^{\rm H}\boldsymbol{\Theta}_k\mathbf{n}_s\vert ^2+\sigma_k^2}, ~\forall k.
	\end{eqnarray} 
	
	\vspace{-2mm}
	\subsection{Problem Formulation}
	In this letter, we aim to maximize the achievable sum rate of all users by jointly optimizing the active beamforming at the BS and the coefficients at the MF-RIS. Under the transmit and amplification power constraints, and the quality-of-service (QoS) requirement of users, the considered optimization problem can be formulated as
	\begin{subequations}
		\setlength{\abovedisplayskip}{3pt}
		\setlength{\belowdisplayskip}{3pt}	
		\label{P0}
		\begin{eqnarray}
			&\!\!\!\!\!\!\!\! \max \limits_{\mathbf{w}_k,\boldsymbol{\Theta}_k} 
			\label{P0-function}
			&\!\!\!\! \sum\nolimits_{k=1}^K R_k  \\
			\label{P0-C-transmit power}
			&\!\!\!\!\!\!\!\! \mathrm{s.t.} &\!\!\!\! \sum\nolimits_{k=1}^{K}\Vert \mathbf{w}_k \Vert ^2 \leq P_{\max},  \\
			\label{P0-C-amplification power}
			&&\!\!\!\! \sum\nolimits_{k=1}^{K}(\Vert \boldsymbol{\Theta}_k\mathbf{H}\mathbf{w}_k \Vert ^2 \!+ \!\Vert \boldsymbol{\Theta}_k\mathbf{I}_M \Vert_F ^2\sigma_s^2) \! \leq \! P_{o}, \\
			\label{P0-C-amplification factor}
			&&\!\!\!\! \beta_m^r+\beta_m^t \le \beta_{\rm{max}},~0 \le \beta_m^p \le \beta_{\max},~\forall m, ~\forall p, \\	
			\label{P0-C-R-min}
			&&\!\!\!\! R_k \ge R_k^{\min}, ~ \theta_m^p \in [0,2\pi), ~(\ref{channel gain}),~\forall k, ~\forall m, ~\forall p,	
		\end{eqnarray}
	\end{subequations}
	where $P_{\rm max}$ and $P_{o}$ denote the maximum transmit and amplification power at the BS and MF-RIS, respectively. $R_k^{\min}$ represents the minimum rate requirement of user $k$. Specifically, the constraints for transmit power, amplification power, the QoS requirements and the decoding order are given in
	(\ref{P0-C-transmit power})-(\ref{P0-C-R-min}),
	respectively. 
	It can be observed that the formulated problem (\ref{P0}) is intractable due to the non-convex objective function and constraints. Besides, the active beamforming and MF-RIS coefficients are highly coupled, making it difficult to be solved directly. Thus, we aim to transform problem (\ref{P0}) into some tractable convex subproblems and solve them separately and alternatively over iterations.
	In the next section, we adopt alternating optimization method to obtain the active beamforming and the MF-RIS coefficients efficiently. 	
	
	\vspace{-3mm}
	\section{Proposed Solution}
	\vspace{-2mm}
	\subsection{Active Beamforming Design}
	Given the MF-RIS coefficients, the active beamforming optimization problem is still non-convex. To solve it, we first introduce an auxiliary variable set $\{A_{k},B_{k}|k \in \mathcal{K}\}$, where $A_k$ and $B_k$ are defined as
	\setlength{\abovedisplayskip}{3pt}
	\setlength{\belowdisplayskip}{3pt}
	\begin{eqnarray}
		\label{A_k}
		&& {A_k}^{-1} =\vert \widehat{\mathbf{h}}_k\mathbf{w}_k \vert^2,   \\
		\label{B_k}
		&& B_k = \sum\nolimits_{i=k+1}^K (\vert \widehat{\mathbf{h}}_{k}\mathbf{w}_{i} \vert ^2)+\vert \mathbf{g}_k^{\rm H}\boldsymbol{\Theta}_k\mathbf{n}_s\vert ^2+\sigma_k^2.
	\end{eqnarray}
	Thus, the achievable data rate can be rewritten as
	$R_k=\log_2 \big(1+{(A_kB_k)}^{-1} \big)$.
	
	Then, the active beamforming optimization problem in (\ref{P0}) can be equivalently expressed as
	\begin{subequations}
		\label{P1}
		\begin{eqnarray}	
			&\!\!\!\!\!\!\!\!\!\!\!\!\! \max \limits_{\mathbf{w}_k,A_k,B_k,R_k} 
			\label{P1-function}
			& \!\!\!\!\! \sum\nolimits_{k=1}^K R_k  \\
			\label{P1-C-R_k}
			&\!\!\!\!\!\!\!\!\!\!\!\!\! \mathrm{s.t.} &\!\!\!\!\! \log_2 \big(1+{(A_kB_k)}^{-1}\big) \ge R_k, ~\forall k, \\
			\label{P1-C-A_k}
			&& \!\!\!\!\! {A_k}^{-1} \! \le \vert \widehat{\mathbf{h}}_k\mathbf{w}_k \vert ^2,~\forall k, \\
			\label{P1-C-B_k}
			&& \!\!\!\!\! B_k \! \ge  \!\!  \sum\limits_{i=k+1}^K  (\vert \widehat{\mathbf{h}}_{k}\mathbf{w}_{i} \vert ^2)\!+\!\vert \mathbf{g}_k^{\rm H}\boldsymbol{\Theta}_k\mathbf{n}_s\vert ^2\!+\!\sigma_k^2,~\forall k, \\
			\label{P1-C-R_kmin}
			&& \!\!\!\!\! R_k \ge R_k^{\min}, ~\mathrm{(\ref{P0-C-transmit power})},~ \mathrm{(\ref{P0-C-amplification power})},~\forall k. 
		\end{eqnarray}
	\end{subequations}
	
	We further define $\widehat{\mathbf{H}}_k=\widehat{\mathbf{h}}_k^{\mathrm{H}}\widehat{\mathbf{h}}_k$, $\mathbf{D}_k=(\mathbf{H}^{\mathrm{H}}\mathbf{\Theta}_k)(\mathbf{H}^{\mathrm{H}}\mathbf{\Theta}_k)^{\mathrm{H}}$ and $\mathbf{W}_k=\mathbf{w}_k\mathbf{w}_k^{\mathrm{H}}$,
	where $\mathbf{W}_k \succeq \mathbf{0}$, and $\rm{rank}(\mathbf{W}_k)=1$. Then, we have 
	\begin{eqnarray}
		~~ \vert \widehat{\mathbf{h}}_{k}\mathbf{w}_{k}\vert ^2 
		=\mathrm{Tr}(\widehat{\mathbf{H}}_k \mathbf{W}_{k}), \
		\Vert \boldsymbol{\Theta}_k\mathbf{H}\mathbf{w}_k \Vert ^2  
		=\mathrm{Tr}(\mathbf{W}_{k} \mathbf{D}_k). 
	\end{eqnarray}
	
	Therefore, problem (\ref{P1}) can be reformulated as
	\begin{subequations}
		\label{P2}
		\begin{eqnarray}
			&\!\!\!\!\!\!\!\!\!\!\!\!\!\!\! \max \limits_{\mathbf{W}_k,A_k,B_k,R_k} 
			\label{P2-function}
			&\!\!\!\!\!\! \sum\nolimits_{k=1}^K R_k  \\ 
			\label{P2-C-A_k}
			&\!\!\!\!\!\!\!\!\!\!\!\!\!\!\! \mathrm{s.t.} & \!\!\!\!\!\! {A_k}^{-1} \le \mathrm{Tr}(\widehat{\mathbf{H}}_k\mathbf{W}_k), ~\forall k,\\
			\label{P2-C-B_k}
			&& \!\!\!\!\!\! B_k \!\! \ge \!\! \sum\limits_{i=k+1}^K \!\! \mathrm{Tr}(\widehat{\mathbf{H}}_k \mathbf{W}_{i})\!+\!\vert \mathbf{g}_k^{\rm H}\boldsymbol{\Theta}_k\mathbf{n}_s\vert ^2\!+\!\sigma_k^2,~\forall k, \\
			\label{P2-C-transmit power}
			&&\!\!\!\!\!\! \sum\nolimits_{k=1}^K \!\mathrm{Tr}(\mathbf{W}_k) \le P_{\mathrm{max}},\\
			\label{P2-C-amplification power}
			&&\!\!\!\!\!\! \sum\nolimits_{k=1}^K \big [\mathrm{Tr}(\mathbf{W}_{k}\mathbf{D}_k)\! + \!\Vert \boldsymbol{\Theta}_k\mathbf{I}_M \Vert ^2 \!\sigma_s^2 \big ] \!\le\! P_{o},  \\
			\label{P2-C-W_k-rank-1}
			&&\!\!\!\!\!\! \mathrm{rank}(\mathbf{W}_k)=1, ~\forall k,
			\\
			\label{P2-C-W_k}
			&&\!\!\!\!\!\! \mathbf{W}_k \succeq \mathbf{0}, ~R_k \ge R_k^{\min},~\mathrm{(\ref{P1-C-R_k})}, ~\forall k.	
		\end{eqnarray}
	\end{subequations}
	
	In order to deal with the non-convex constraint (\ref{P1-C-R_k}), we adopt the first-order Taylor expansion, and then we obtain the lower bound as follows:
	\begin{align}
		\setlength{\abovedisplayskip}{3pt}
		\setlength{\belowdisplayskip}{3pt}	
		\label{A_k-B_k_taylor expansion}
		\log_2(1 \!+\! \frac{1}{A_kB_k})\!  & \ge \! \log_2(1\! +\! \frac{1}{A_k^{(\tau_1)}B_k^{(\tau_1)}}) 
		\!\!-\!\! \frac{\log_2e(A_k \!-\! A_k^{(\tau_1)})}{A_k^{(\tau_1)}(1 \!+\! A_k^{(\tau_1)}B_k^{(\tau_1)})}  \nonumber \\
		&-\! \frac{\log_2e(B_k \!\! -\!\! B_k^{(\tau_1)})}{B_k^{(\tau_1)}(1 \!\!+\!\! A_k^{(\tau_1)}B_k^{(\tau_1)})} \stackrel{\Delta}{=} \overline{R}_k,    
	\end{align}
	where $A_k^{(\tau_1)}$ and $B_k^{(\tau_1)}$ are feasible points of $A_k$ and $B_k$ in the $\tau_1$-th iteration, respectively.
	
	For the non-convex rank-one constraint in (\ref{P2-C-W_k-rank-1}), we assume to transform it to a penalty term in the objective function, which can be solved by SCA. Thus, we firstly introduce an equivalent equality:
	\begin{eqnarray}
		\label{W_k-rank-1-one}
		\Vert \mathbf{W}_k \Vert_{\ast}-\Vert \mathbf{W}_k \Vert_2=0,~\forall k,
	\end{eqnarray}
	where $\Vert \mathbf{W}_k \Vert_{\ast}=\sum_{i}\varepsilon_i(\mathbf{W}_k)$ and $\Vert \mathbf{W}_k \Vert_2=\varepsilon_1(\mathbf{W}_k)$ denote the nuclear norm and the spectral norm of $\mathbf{W}_k$, respectively. $\varepsilon_i(\mathbf{W}_k)$ is the $i$-th largest singular value of matrix $\mathbf{W}_k$. Thus, when the matrix $\mathbf{W}_k$ is rank-one, equality (\ref{W_k-rank-1-one}) holds. 
	
	Next, we employ the penalty method to solve problem (\ref{P2}) by adding (\ref{W_k-rank-1-one}) to the objective function (\ref{P2-function}). Since the penalty term (\ref{W_k-rank-1-one}) makes the objective function not convex, we apply the first-order Taylor expansion to obtain a convex upper bound of (\ref{W_k-rank-1-one}) as follows:
	\begin{align}
		\label{W_k-rank-1-two}
		\Vert \mathbf{W}_k \Vert_{\ast}-\Vert \mathbf{W}_k \Vert_2   \le 
		\Vert \mathbf{W}_k \Vert_{\ast}- \Vert \overline{\mathbf{W}}_k \Vert_2,
	\end{align}
	where $\Vert \overline{\mathbf{W}}_k \Vert_2=\Vert \mathbf{W}_k^{(\tau_1)}  \Vert_2+\mathrm{Tr} \big [\mathbf{e}_k^{(\tau_1)}(\mathbf{e}_k^{(\tau_1)})^{\mathrm{H}}(\mathbf{W}_k-\mathbf{W}_k^{(\tau_1)}) \big]$, and $\mathbf{e}_k^{(\tau_1)}$ is the eigenvector corresponding to the largest eigenvalue of $\mathbf{W}_k^{(\tau_1)}$ in the $\tau_1$-th iteration. 
	
	By introducing (\ref{W_k-rank-1-two}) to the objective function (\ref{P2-function}), we obtain the following problem:
	\begin{subequations}
		\setlength{\abovedisplayskip}{3pt}
		\setlength{\belowdisplayskip}{3pt}	
		\label{P3}
		\begin{eqnarray}
			&\!\!\!\!\!\!\!\!\!\! \max \limits_{\mathbf{W}_k,A_k,B_k,R_k} 
			\label{P3-function}
			& \!\!\!\!\!\! \sum\nolimits_{k=1}^K \! R_k \! - \! \frac{1}{\eta}\sum\nolimits_{k} \!(\Vert \mathbf{W}_k \Vert_{\ast} \!\!- \!\!\Vert \overline{\mathbf{W}}_k \Vert _2) \\ 
			\label{P3-C-R_k}
			&\!\!\!\!\!\!\!\!\!\! \mathrm{s.t.} &\!\!\!\!\!\! \overline{R}_k \ge R_k,~\mathbf{W}_k \succeq \mathbf{0}, ~R_k \ge R_k^{\min}, ~\forall k,\\
			\label{P3-others}
			&& \!\!\!\!\!\!  \mathrm{(\ref{P2-C-A_k})} - \mathrm{(\ref{P2-C-amplification power})},
		\end{eqnarray}
	\end{subequations}
	where $\eta \textgreater 0$ is the penalty factor penalizing (\ref{P3-function}) if $\mathbf{W}_k$ is not rank-one. 
	It can be verified that, when $\eta \rightarrow 0$, the solution $\{\mathbf{W}_k\}$ of problem (\ref{P3}) always satisfies equality (\ref{W_k-rank-1-one}). 
	
	The reformulated problem (\ref{P3}) is a standard convex semi-definite programming (SDP), which can be efficiently solved via CVX. 
	To obtain a high quality solution, we first initialize a large $\eta$ to find a feasible starting point, and then gradually decrease $\eta$ with $\eta=\mu\eta,\mu \textless 1$ to a sufficiently small value to obtain an overall suboptimal solution. 
	The process terminates when the penalty term satisfies the following criterion:
	\begin{eqnarray}
		\label{W-rank1-two}
		\max\{\Vert \mathbf{W}_k \Vert_{\ast}-\Vert \overline{\mathbf{W}}_k \Vert _2,~\forall k \} \le \epsilon_1,
	\end{eqnarray}
	where $\epsilon_1$ denotes a predefined maximum violation of (\ref{W_k-rank-1-one}). 
	
	\vspace{-3.5mm}
	\subsection{MF-RIS Coefficient Design}
	For the coefficient design at the MF-RIS, we define $\mathbf{v}_k =  [\mathbf{u}_r;1]$ if user $k$ is located at the space $r$; otherwise $\mathbf{v}_k =  [\mathbf{u}_t;1]$. Then, we define $\mathbf{V}_k = \mathbf{v}_k\mathbf{v}_k^{\mathrm{H}}$, with $\mathbf{V}_k  \succeq  \mathbf{0}$ and $\mathrm{rank}(\mathbf{V}_k) = 1$. Let $\mathbf{g}_k = [g_{k,1},g_{k,2}, \ldots, g_{k,M}]^{\mathrm{H}}$ and $\mathbf{G}_k=\mathbf{H}\mathbf{w}_k$, then we have $\mathbf{Q}_k = \mathrm{diag}\big (\big [\vert g_{k,1} \vert ^2,\vert g_{k,2} \vert ^2,\ldots,\vert g_{k,M} \vert ^2 \big ] \big)$ and  ${\widehat{\mathbf{G}}}_k=\mathrm{diag} \big (\big [\vert \mathbf{G}_{k,1} \vert^2,\vert \mathbf{G}_{k,2} \vert^2,\ldots,\vert \mathbf{G}_{k,M} \vert^2 \big] \big )+  \sigma_s^2\mathbf{I}_M $. Given 
	\begin{eqnarray}
		\mathbf{\overline{Q}}_k=\left[
		\begin{array}{cc}
			\mathbf{Q}_k & \mathbf{0} \\
			\mathbf{0} & 0
		\end{array}
		\right],
		\overline{\mathbf{G}}_k=\left[
		\begin{array}{cc}
			{\widehat{\mathbf{G}}}_k & \mathbf{0} \\
			\mathbf{0} & 0
		\end{array}
		\right], 
	\end{eqnarray}
	we can obtain
	\begin{eqnarray}
		\label{amplification power budget}
		&& \Vert \boldsymbol{\Theta}_k\mathbf{H}\mathbf{w}_{k} \Vert ^2+ \Vert \boldsymbol{\Theta}_k\mathbf{I}_M \Vert_F ^2\sigma_s^2= \mathrm{Tr}(\mathbf{V}_k\mathbf{\overline{G}_k)},\\
		\label{}
		&& \Vert \mathbf{g}_k^\mathrm{H} \boldsymbol{\Theta}_k \Vert^2 = \mathrm{Tr}(\mathbf{V}_k\mathbf{\overline{Q}}_k).
	\end{eqnarray}
	Thus, constraint (\ref{P0-C-amplification power}) can be replaced by (\ref{amplification power budget}).
	
	In order to handle the non-convex constraints (\ref{channel gain}) and (\ref{A_k}), we define
	$\mathbf{f}_k=\mathrm{diag}(\mathbf{g}_k^{\mathrm{H}})\mathbf{G}_k$, $\mathbf{R}_k=\mathrm{diag}(\mathbf{g}_k^{\mathrm{H}})\mathbf{H}$, $\tilde{h}_k=\Vert \mathbf{h}_k^{\mathrm{H}} \Vert ^2$, and $d_k =\mathbf{w}_k^{\mathrm{H}}\mathbf{h}_k $, then we have 
	\begin{eqnarray}
		\mathbf{F}_k=\left[
		\begin{array}{cc}
			\mathbf{f}_k\mathbf{f}_k^{\mathrm{H}} & \mathbf{f}_kd_k^{\ast} \\
			d_k\mathbf{f}_k^{\mathrm{H}} & \vert d_k \vert^2 
		\end{array}
		\right], 
		\mathbf{\overline{R}}_k=\left[
		\begin{array}{cc}
			\mathbf{R}_k\mathbf{R}_k^{\mathrm{H}} & \mathbf{R}_k\mathbf{h}_k \\
			\mathbf{h}_k^{\mathrm{H}}\mathbf{R}_k^{\mathrm{H}} & \tilde{h}_k  
		\end{array}
		\right]. 
	\end{eqnarray}
	According to the above transformation, we can obtain 
	\begin{eqnarray}
		&&\vert \widehat{\mathbf{h}}_k\mathbf{w}_k \vert^2 =\vert (\mathbf{h}_k^{\mathrm{H}}+\mathbf{g}_k^{\mathrm{H}}\boldsymbol{\Theta}_k\mathbf{H})\mathbf{w}_k \vert^2 
		=\mathrm{Tr}(\mathbf{V}_k\mathbf{F}_k), \\ 
		\label{SIC_transformation}
		&& \Vert \widehat{\mathbf{h}}_k  \Vert ^2 = \mathrm{Tr}(\mathbf{V}_k\mathbf{\overline{R}}_k).
	\end{eqnarray}
	
	Based on (\ref{SIC_transformation}), the decoding order in (\ref{channel gain}) is rewritten as
	\begin{eqnarray}
		\label{SIC-solve}
		&& \mathrm{Tr}(\mathbf{V}_1\mathbf{\overline{R}}_1) \le \mathrm{Tr}(\mathbf{V}_2\mathbf{\overline{R}}_2) \le \cdots \le \mathrm{Tr}(\mathbf{V}_K\mathbf{\overline{R}}_K).
	\end{eqnarray}
	
	Then, given the active beamforming vector, the subproblem of MF-RIS coefficient design can be given by
	\begin{subequations}
		\label{P4}
		\begin{eqnarray}
			& \!\!\!\!\!\!\!\!\!\!\!\!\!\!\!\!\!\! \max \limits_{\mathbf{V}_k,A_k,B_k,R_k}
			\label{P4-function}
			& \!\!\!\!\!\! \sum\nolimits_{k=1}^K R_k \\
			\label{P4-C-A_k}
			& \!\!\!\!\!\!\!\!\!\!\!\!\!\!\!\!\!\! \mathrm{s.t.} & \!\!\!\!\!\! 
			{A_k}^{-1} \! \le \! \mathrm{Tr}(\mathbf{V}_k\mathbf{F}_{k}), ~\forall k,  \\
			\label{P4-C-B_k}
			&& \!\!\!\!\!\!\! B_k  \!\! \le \!\! \sum\limits_{i=k+1}^K  \!\! \mathrm{Tr}(\!\mathbf{V}_k\mathbf{F}_{i}\!)\!+\! \sigma_s^2\mathrm{Tr}(\!\mathbf{V}_k\mathbf{\overline{Q}}_k\!) \! + \!\sigma_k^2, ~\forall k,\\
			\label{P4-C-amplification power}
			&& \!\!\!\!\!\!\! \sum\nolimits_{k=1}^K \mathrm{Tr}(\mathbf{V}_k{\overline{\mathbf{G}}}_k) \le P_{o},\\
			\label{P4-V_k}
			&&\!\!\!\!\!\!\! \mathbf{V}_k  \succeq  \mathbf{0}, ~R_k  \ge  R_k^{\min},  ~  \forall k,  \\
			\label{P4-V_k-rank-m}
			&& \!\!\!\!\!\!\! [\mathbf{V}_k]_{\rm{m,m}}=\beta_m^k,~[\mathbf{V}_k]_{\rm{M+1,M+1}}=1,~\forall k,\\
			\label{P4-V_k-rank-1}
			&& \!\!\!\!\!\!\! \mathrm{rank}(\mathbf{V}_k)=1, ~\forall k, \\
			\label{P4-other constraints}
			&& \!\!\!\!\!\!\!
			\theta_m^p \in [0,2\pi),~\mathrm{(\ref{P0-C-amplification factor})}, ~\mathrm{(\ref{P1-C-R_k})}, ~  \mathrm{(\ref{SIC-solve})}, ~\forall m,  
			~ \forall p,
		\end{eqnarray}
	\end{subequations}	
	where $\mathbf{F}_{i}$ denotes $\mathbf{F}_k$ when $\mathbf{w}_k$ is replaced by $\mathbf{w}_{i}$.
	
	Similar to (\ref{W_k-rank-1-two}), we replace the rank-one constraint in (\ref{P4-V_k-rank-1}) with the following form:
	\begin{eqnarray}
		\Vert \mathbf{V}_k \Vert_{\ast} - \Vert \mathbf{V}_k \Vert_2   \le 
		\Vert \mathbf{V}_k \Vert_{\ast}- \Vert \overline{\mathbf{V}}_k \Vert_2,
	\end{eqnarray}
	where $\Vert \mathbf{V}_k \Vert_{\ast}$ and $\Vert \mathbf{V}_k \Vert_2$ denote the nuclear norm and the spectral norm of matrix $\mathbf{V}_k$, respectively. Besides,
	$\Vert \overline{\mathbf{V}}_k \Vert_2=\Vert \mathbf{V}_k^{(\tau_2)}  \Vert_2+\mathrm{Tr} \big[\mathbf{z}_k^{(\tau_2)}(\mathbf{z}_k^{(\tau_2)})^{\mathrm{H}}(\mathbf{V}_k-\mathbf{V}_k^{(\tau_2)}) \big]$ and $\mathbf{z}_k^{(\tau_2)}$ is the eigenvector corresponding to the largest eigenvalue of $\mathbf{V}_k^{(\tau_2)}$ in the $\tau_2$-th iteration.
	
	By introducing (\ref{A_k-B_k_taylor expansion}) into (\ref{P1-C-R_k}), problem (\ref{P4}) can be reformulated as
	\begin{subequations}
		\setlength{\abovedisplayskip}{3pt}
		\setlength{\belowdisplayskip}{3pt}		
		\label{P5}
		\begin{eqnarray}
			&\!\!\!\!\!\!\!\!\!\! \max \limits_{\mathbf{V}_k,A_k,B_k,R_k}
			\label{P5-function}
			& \!\!\!\!\! \sum\nolimits_{k=1}^K \! R_k \! -\! \frac{1}{\xi}\sum\nolimits_{k} \! (\Vert \mathbf{V}_k \Vert_{\ast} \!- \!\Vert \overline{\mathbf{V}}_k \Vert_2) \\
			\label{P5-all}
			&\!\!\!\!\!\!\!\!\!\!  \mathrm{s.t.} & \!\!\!\!\!	
			\overline{R}_k \ge R_k,  ~\theta_m^p \in [0,2\pi), ~\forall k,~\forall m,~\forall p,\\
			&& \!\!\!\!\! \mathrm{(\ref{P0-C-amplification factor})},~\mathrm{(\ref{SIC-solve})}, 
			~\mathrm{(\ref{P4-C-A_k})}\!\!- \!\! \mathrm{(\ref{P4-V_k-rank-m})},	
		\end{eqnarray}
	\end{subequations}
	where $\xi \textgreater 0$ is the penalty factor to ensure $\mathbf{V}_k$ is rank-one.
	
	The problem (\ref{P5}) is a standard SDP problem. It can be solved by CVX. The termination criterion is given by
	\setlength{\abovedisplayskip}{3pt}
	\setlength{\belowdisplayskip}{3pt}
	\begin{eqnarray}
		\label{V-rank1-two}
		\max\{\Vert \mathbf{V}_k \Vert_{\ast}-\Vert \overline{\mathbf{V}}_k \Vert _2,~\forall k \} \le \epsilon_2,
	\end{eqnarray}
	where $\epsilon_2$ denotes a predefined maximum violation. 
	
	\begin{algorithm}[t]  
		\caption{Penalty-Based Iterative Algorithm}
		\label{Algorithm2}
		\renewcommand{\algorithmicrequire}{\textbf{Initialize}}
		\renewcommand{\algorithmicensure}{\textbf{Output}}
		\begin{algorithmic}[1]	
			\STATE \textbf{Initialize} $\{\mathbf{W}_k^{(0)}\}, \{\mathbf{V}_k^{(0)}\}$, the error tolerance $\Delta$, the maximum number of iteration $T_{0,\max}$, the penalty factors $\eta$ and $\xi$, and the predefined threshold $\epsilon$.
			\REPEAT 
			\STATE Set the iteration index $\tau_0=0$;
			\REPEAT
			\STATE Given $\mathbf{V}_k^{(\tau_0)}$, update $\mathbf{W}_k^{(\tau_0+1)}$ by solving (\ref{P3});
			\STATE Given $\mathbf{W}_k^{(\tau_0+1)}$, update $\mathbf{V}_k^{(\tau_0+1)}$ by solving (\ref{P5});
			\STATE Update $\tau_0=\tau_0+1$;
			\UNTIL $\vert \frac{R_{\mathrm{sum}}^{(\tau_0)}-R_{\mathrm{sum}}^{(\tau_0-1)}}{R_{\mathrm{sum}}^{(\tau_0-1)}} \vert \textless \Delta $ or $\tau_0 \textgreater T_{0,\max}$.
			\STATE Update \{$\mathbf{W}_k^{(0)}, \mathbf{V}_k^{(0)}$\} with  \{$\mathbf{W}_k^{(\tau_0)}, \mathbf{V}_k^{(\tau_0)}$\};
			\STATE Update $\eta=\mu\eta, \xi=\mu\xi $;
			\UNTIL the constraints (\ref{W-rank1-two}) and (\ref{V-rank1-two}) satisfy $\epsilon$;
			\STATE \textbf{Output} the converged solutions $\{\mathbf{W}_k^{\ast}\}$ and $\{\mathbf{V}_k^{\ast}\}$.	
		\end{algorithmic}
	\end{algorithm}	
	
	Based on the above derivation, we propose a penalty-based iterative algorithm to solve problem (\ref{P0}) efficiently. The details are given in Algorithm \ref{Algorithm2}. 
	Specifically, the initial points $\{\mathbf{W}_k^{(0)}\}$ and $\{\mathbf{V}_k^{(0)}\}$ are obtained by selecting the feasible ones from some random points.
	Since both the objectives of problems (\ref{P3}) and (\ref{P5}) are non-decreasing over iterations and the system throughout is upper-bounded by a finite value, the proposed Algorithm \ref{Algorithm2} is guaranteed to converge. 
	Moreover, if the interior point method is employed, the complexity of Algorithm \ref{Algorithm2} is $\mathcal{O}(I_{\mathrm{out}}I_{\mathrm{in}}(KN^{3.5}+2M^{3.5}))$, where $K$, $M$ and $N$ are the numbers of users, BS antennas and MF-RIS elements, respectively. 
	The terms $I_{\mathrm{in}}$ and $I_{\mathrm{out}}$ denote the number of the inner and outer iterations required for convergence, respectively.
	
	\begin{table}[t]  	
		\centering
		\renewcommand{\arraystretch}{1.25}
		\caption{Simulation Parameters}
		\label{tab1}
		\vspace{-0.1cm}
		\label{parameter}
		\scalebox{1}{   
			\begin{tabular}{|l|l|}
				\hline  
				\bfseries Parameter & \bfseries Value\\
				\hline
				\tabincell{l}
				{Path loss exponents of BS-MF-RIS, \\ 
					BS-users, MF-RIS-users links} & 2.5, 3.5, 2.8 \\
				\hline
				\tabincell{l}
				{Rician factors of all links}  & $3~\mathrm{dB}$ \\
				\hline
				Noise power at MF-RIS and users & $- 80~\mathrm{dBm}$  \\ 
				\hline
				Minimum required QoS for users & $0.1~\mathrm{bit/s/Hz}$\\
				\hline
				Maximum amplification power \cite{MF-RISpowerbudgetLong} & $P_{\mathrm{o}}=10~\mathrm{dBm}$ \\
				\hline
				Maximum amplification factor  & $\beta_{\mathrm{max}}=22~\mathrm{dB}$ \\  
				\hline
				Convergence tolerance & $\Delta=10^{-6}$ \\
				\hline
		\end{tabular}}
		\vspace{-5mm}
	\end{table}
	
	\section{Simulation Results}
	In this section, numerical results are provided to validate the performance of an MF-RIS-aided NOMA network.
	The BS and the MF-RIS are located at $(0,0,0)$ and $(0,50,20)$, respectively.
	Besides, the users are divided into two parts, distributed on circles centered on $(0, 45, 0)$ and $(0, 55, 0)$ with radius $r=3$, respectively.
	We adopt Rician fading for all channels, and set $K = 6, ~N = 16, ~M = 100$, and $P_{\max} = 20 ~\mathrm{dBm}$. 
	Other parameters are listed in Table \ref{parameter}. We compare the proposed MF-RIS with three existing RISs:
	
	\begin{itemize}	
		\item
		SF-RIS \cite{Ni2021RIS}: The SF-RIS only supports signal reflection or transmission, i.e., $\beta_{\max}=1$  and $\boldsymbol{\Theta}_t$ or $\boldsymbol{\Theta}_r$ $=\mathbf{0}_{M\times M}$.
		
		\item
		Active RIS \cite{MF-RISpowerbudgetLong}: The active RIS simultaneously supports signal reflection and amplification, i.e., $\boldsymbol{\Theta}_t=\mathbf{0}_{M\times M}$.
		
		\item
		STAR-RIS \cite{Wu2021STARRIS}: The STAR-RIS provides full space coverage by splitting signals to two sides, i.e., $\beta_{\max}=1$.
	\end{itemize}
	
	\begin{figure*}[t]
		\centering
		\subfigure[Sum rate vs. the power budget.]{
			\begin{minipage}[t]{0.31\linewidth}
				\centering
				\includegraphics[width=2.3in]{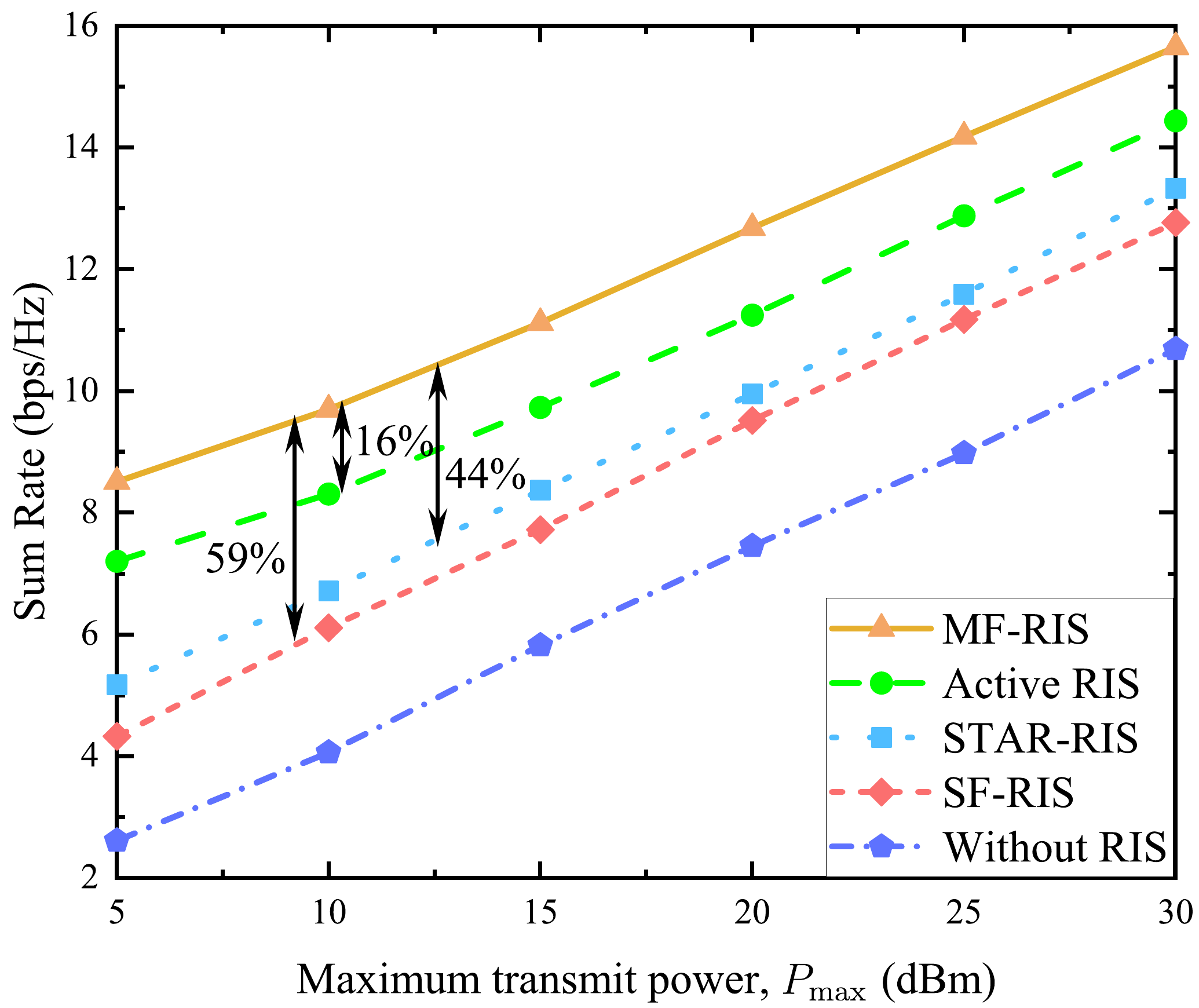}
				\label{maximum transmit power in BS}
				\vspace{-25 pt}
			\end{minipage}
		}
		\subfigure[Sum rate vs. the number of elements.]{
			\begin{minipage}[t]{0.31\linewidth}
				\centering
				\includegraphics[width=2.3in]{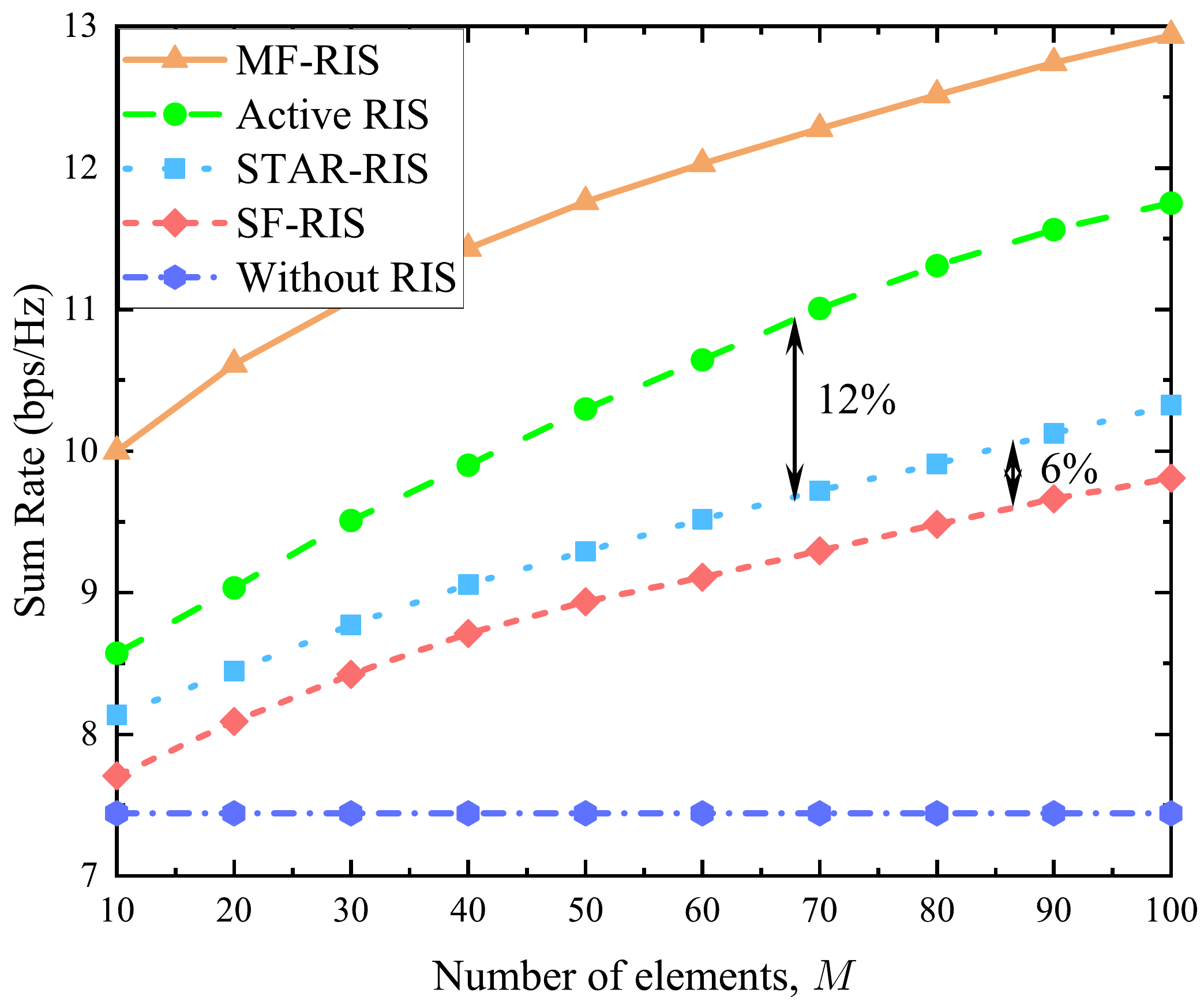}
				\label{Number of elements M}
				\vspace{-25 pt}
			\end{minipage}
		}
		\subfigure[Sum rate vs. the $Y$-coordinate of RIS.]{
			\begin{minipage}[t]{0.31\linewidth}
				\centering
				\includegraphics[width=2.3in]{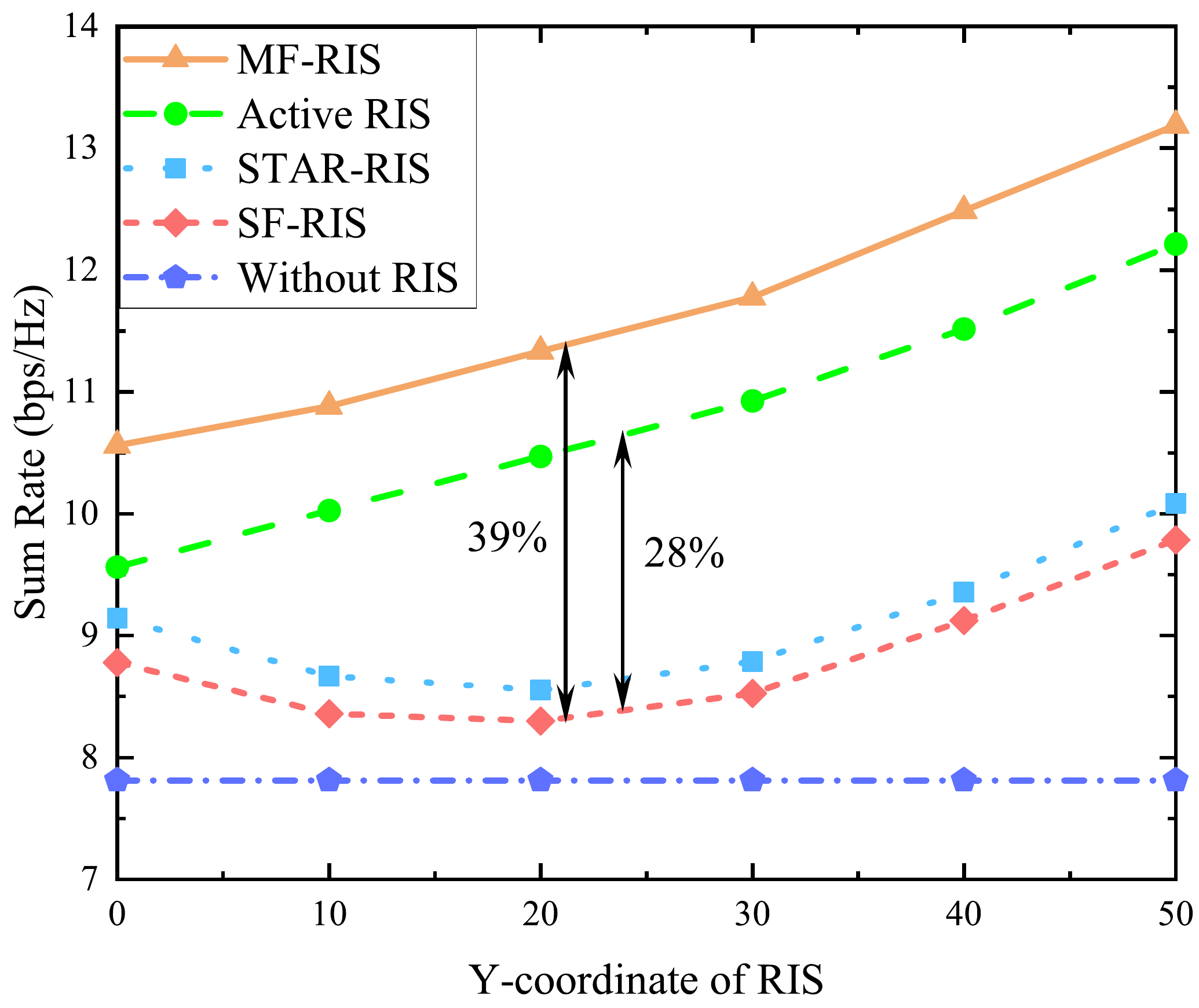}
				\label{Distance between BS and RIS}
				\vspace{-25 pt}
			\end{minipage}
		}
		\vspace{-10 pt}
		\centering
		\caption{Simulation results for the sum rate versus different transmit power, number of elements and RIS locations.}
		\vspace{-15 pt}
	\end{figure*}
	
	Fig. \ref{maximum transmit power in BS} depicts the sum rate versus the maximum transmit power $P_{\max}$. 
	It can be observed that the sum rates of all schemes increase with $P_{\max}$. Besides, the proposed MF-RIS always yields a better performance than other benchmarks. Specifically, when $P_{\max}=10~\mathrm{dBm}$, the MF-RIS enjoys 
	a 59\% higher sum rate than the SF-RIS.
	This is because the MF-RIS serves all users in full space through signal reflection, transmission and amplification functions.
	Besides, by providing additional energy to amplify the incident signal, the MF-RIS is able to efficiently mitigate the double-fading attenuation, which helps to improve the channel gain of cascaded links. Furthermore, due to the limitations faced by the active RIS and STAR-RIS counterparts (i.e., half-space coverage and double-fading attenuation), 
	the MF-RIS improves the rate performance by 16\% and 44\% when $P_{\max}=10~\mathrm{dBm}$, respectively. 
	Additionally, it is evident that all RIS-aided schemes achieve significant gains than the scheme without RIS.
	This demonstrates the superiority of using RIS to improve the performance of wireless networks.	
	
	Fig. \ref{Number of elements M} shows that the sum rate of all RIS-aided schemes increase with $M$. 	
	This is because a larger $M$ enables a higher beamforming gain, thus improving the system performance.
	In addition, with more degree of freedoms to manipulate signal propagation, the STAR-RIS is capable of enjoying a 6\% higher sum rate than the SF-RIS.
	Moreover, although only the users located in the reflection space are served by the active RIS, it outperforms the STAR-RIS with a 12\% higher sum rate. This is because the performance gain obtained from the signal amplification of active RIS is greater than that from full-space coverage of STAR-RIS. This also implies that the signal amplification function plays an important role in improving the performance of RIS-aided networks.
	
	Fig. \ref{Distance between BS and RIS} illustrates the sum rate versus the $Y$-coordinate of RIS (from $0$ to 50), where the RIS moves from the BS side to the user side.
	We can observe that the sum rates of the STAR-RIS and the SF-RIS first decrease and then increase. The reason behind this is that the channel gain decreases with the link distance. 
	Specifically, when the STAR-RIS and the SF-RIS are located close to the middle point, the received signals at users are attenuated the most, resulting in the lowest sum rate. In contrast, owing to the signal amplification function, the MF-RIS and the active RIS are less affected by the double-fading attenuation, which achieve 39\% and 28\% gains in the middle point compared to the SF-RIS. 
	Moreover, the corresponding sum rate maintains a continuous upward trend even when the MF-RIS and active RIS are far away from the BS. This is because as the RIS comes closer to the users, the power of the incident signal at the RIS is weaker. Thus, under a fixed amplification power budget, the MF-RIS can provide more amplification gain when deployed closer to users. This compensates for the attenuation caused by the double-fading issue.
	This observation also reveals that the MF-RIS should be deployed close to the users for better performance. 
	
	\vspace{-4mm}
	\section{Conclusion} \label{Conclusion}   
	In this letter, we proposed a novel MF-RIS architecture to alleviate the double-fading attenuation via transmitting and reflecting the incident signal with power amplification. Then, we investigated the resource allocation problem in a downlink multiuser MF-RIS-aided NOMA network. 
	Specifically, the active beamforming and MF-RIS coefficients were jointly optimized to maximize the achievable sum rate by leveraging SCA and penalty-based method. Numerical results validated the effectiveness of the proposed MF-RIS and the superiority of MF-RIS over traditional RISs. In the future, we are interested in studying the coupled phase and hardware impairment problems of the MF-RIS. In addition, the robust beamforming under imperfect CSI cases deserves exploration as well.


\vspace{-4mm}	
	\begin{thebibliography}{99}
		\bibliographystyle{IEEEtran}
		\bibitem{LiuNOMA2017}
		Y.~Liu, Z.~Qin, Elkashlan \emph{et~al.}, ``Nonorthogonal multiple access for {5G} and beyond,'' \emph{Proc. IEEE}, vol. 105, no.~12, pp. 2347--2381, Dec.2017.
		
		\bibitem{ElhattabNOMA}
		M.~Elhattab, M.~A. Arfaoui, C.~Assi \emph{et~al.}, ``{RIS}-assisted joint transmission in a two-cell downlink {NOMA} cellular system,'' \emph{IEEE J. Sel. Areas Commun.}, vol.~40, no.~4, pp. 1270--1286, Apr. 2022.
		
		\bibitem{Liu2021RIS}
		Y.~Liu, X.~Liu, X.~Mu \emph{et~al.}, ``Reconfigurable intelligent surfaces: Principles and opportunities,'' \emph{IEEE Commun. Surveys Tuts.}, vol.~23, no.~3, pp. 1546--1577, thirdquarter 2021.
		
		\bibitem{SenaRIS-NOMA2020}
		A.~S.~d. Sena, D.~Carrillo, F.~Fang \emph{et~al.}, ``What role do intelligent reflecting surfaces play in multi-antenna non-orthogonal multiple access?'' \emph{IEEE Wireless Commun.}, vol.~27, no.~5, pp. 24--31, Oct. 2020.
		
		\bibitem{DingRIS-NOMA2020}
		Z.~Ding and H.~Vincent~Poor, ``A simple design of {IRS-NOMA} transmission,'' \emph{IEEE Commun. Lett.}, vol.~24, no.~5, pp. 1119--1123, Feb. 2020.
		
		\bibitem{ZhengRIS-NOMA2020}
		B.~Zheng, Q.~Wu, and R.~Zhang, ``Intelligent reflecting surface-assisted multiple access with user pairing: {NOMA} or {OMA}?'' \emph{IEEE Commun. Lett.}, vol.~24, no.~4, pp. 753--757, Jan. 2020.
		
		\bibitem{Ni2021RIS}
		W.~Ni, X.~Liu, Y.~Liu \emph{et~al.}, ``Resource allocation for multi-cell {IRS}-aided {NOMA} networks,'' \emph{IEEE Trans. Wireless Commun.}, vol.~20, no.~7, pp. 4253--4268, Jul. 2021.
		
		\bibitem{Wang2021STARRIS}
		W.~Wang, W.~Ni, H.~Tian \emph{et~al.}, ``Safeguarding {NOMA} networks via reconfigurable dual-functional surface under imperfect {CSI},'' \emph{IEEE J. Sel. Topics Signal Process.}, vol.~16, no.~5, pp. 950--966, Aug. 2022.
		
		\bibitem{Wu2021STARRIS}
		C.~Wu, Y.~Liu, X.~Mu \emph{et~al.}, ``Coverage characterization of {STAR-RIS} networks: {NOMA} and {OMA},'' \emph{IEEE Commun. Lett.}, vol.~25, no.~9, pp.3036--3040, Sept. 2021.
		
		\bibitem{WangIOSrate}
		T.~Wang, M.-A. Badiu, G.~Chen \emph{et~al.}, ``Performance analysis of {IOS}-assisted {NOMA} system with channel correlation and phase errors,'' \emph{IEEE Trans. Veh. Technol.}, vol.~71, no.~11, pp. 11\,861--11\,875, Nov. 2022.
		
		\bibitem{Liu2022WCL}
		H.~Liu, G.~Li, X.~Li \emph{et~al.}, ``Effective capacity analysis of {STAR-RIS}-assisted {NOMA} networks,'' \emph{IEEE Wireless Commun. Lett.}, vol.~11, no.~9, pp. 1930--1934, Sept. 2022.
		
		\bibitem{Li2022TVT}
		X.~Li, Y.~Zheng, M.~Zeng \emph{et~al.}, ``Enhancing secrecy performance for {STAR-RIS NOMA} networks,'' \emph{IEEE Trans. Veh. Technol.}, Oct. 2022.
		
		\bibitem{MF-RISpowerbudgetLong}
		R.~Long, Y.-C. Liang, Y.~Pei \emph{et~al.}, ``Active reconfigurable intelligent surface-aided wireless communications,'' \emph{IEEE Trans. Wireless Commun.}, vol.~20, no.~8, pp. 4962--4975, Aug. 2021.
		
	\end{thebibliography}
\end{document}